\documentstyle[twoside,fleqn,npb,epsfig]{article}
\newcommand{\AmS}{{\protect\the\textfont2
  A\kern-.1667em\lower.5ex\hbox{M}\kern-.125emS}}
\newcommand{\pom}{I\!\!P}                
\newcommand{\xpom}{{x_{I\!\!P}}}
\newcommand{\ptsq}{{p_{\bot}^{2}}}
\newcommand{\rmd}{{{\rm d}}}
\newcommand{\ptsqmin}{{p_{\bot\,{\rm min}}^{2}}}
\def\lsim{\mathrel{\rlap{\lower4pt\hbox{\hskip1pt$\sim$}}
    \raise1pt\hbox{$<$}}}                

\title{Analysis of Rapidity Gap Cuts in Diffractive DIS}

\author{J. Williams\address{Theoretical Physics,
        Oxford University, \\
        1 Keble Road, \\
        Oxford, OX1 3NP, U.K.}%
        \thanks{Funded partly by a Commonwealth
        Scholarship, and also by an NZFUW Fellowship.}}

\begin{document}

\begin{abstract}
The requirement of a large pseudo-rapidity gap to select diffractive
DIS events at HERA restricts the kinematically accessible region of phase
space for a significant range of $Q^2$, $\beta$ and
$\xpom$. Consequences of this include a breakdown of
$\xpom$-factorization in large rapidity gap diffractive samples and an
enhancement in the relative contribution of quark-antiquark-gluon
processes over dijet processes in the diffractive DIS sample. 
\end{abstract}

\maketitle

\section{Introduction}

Large Rapidity Gap (LRG) cuts, used in some analyses of diffractive
DIS at HERA to select diffractive events, restrict the kinematically
accessible phase space for a significant range of the kinematic
parameters, $Q^2$, $\beta$ and $\xpom$\,\cite{Ellis:1996cg}. One
consequence of this is a systematic reduction in the diffractive
structure function. Further, since LRG cuts provide a stronger
constraint on diffractive dijet production than on higher-multiplicity
diffractive final states, these cuts are expected to lead to a
relative enhancement in the contribution from $q\bar{q}g$ and
higher-order diffractive final states. The sensitivity of the phase
space constraints to $\xpom$ also means that one cannot extract a
well-defined pomeron structure function from LRG data for which phase
space effects are expected.

In the next section we briefly discuss the relevant kinematics and
hadronization effects. In section\,\ref{sect:consequences} we explain
the relationship between pseudo-rapidity cuts and phase space
constraints in diffractive DIS, and describe the region of HERA
parameter space in which such effects might be observed. Following
this, we explore the consequences of LRG cuts on the extraction of
diffractive and pomeron structure functions, and constraints on
$q\bar{q}g$ and higher-order diffractive final states.

\section{Kinematics} \label{sect:kinematics}

We use the usual variables of DIS and diffractive DIS, where
pseudo-rapidity, $\eta$, is defined by

\begin{equation}
\eta=-\ln\tan\frac{\theta_{\rm lab}}{2},
\end{equation}

\noindent
where $\theta_{\rm lab}$ is the HERA LAB angle between the forward
proton direction and any significant hadronic activity from the
diffractive final state.

Another interesting kinematic variable is the transverse momentum,
$\ptsq$, of final-state partons in the virtual photon-pomeron
centre-of-momentum (CMS) system. For dijet final states, this is given
by

\begin{equation}
\ptsq=\frac{M_X^2}{4}\sin^2\theta_{\rm cms},
\end{equation}

\noindent
where $\theta_{\rm cms}$ is the CMS scattering angle between the final
state parton in the diffractive system which couples to the same
vertex as the pomeron, and the $\gamma^{*}-\pom$ axis. A similar
relation can easily be constructed for higher-order diffractive final
states.

In order to relate pseudo-rapidity cuts, which are made at hadron
level in the HERA LAB frame, to $\ptsq$, defined above at parton level
in the $\gamma^{*}-\pom$ CMS system, one must first make some
assumption about hadronization effects, and also calculate the boost
between the LAB and CMS systems. We assume that the final-state
partons hadronize into a jet with a radius of half a unit of
pseudo-rapidity. Thus, for example, a pseudo-rapidity cut, $\eta_{\rm
max}$, of 3.2 made at hadron level corresponds to a pseudo-rapidity
interval of about 2.7 at parton level.

\section{Consequences of LRG Cuts} \label{sect:consequences}

\subsection{Constraints on Diffractive Final-State Phase Space for
Dijet Production}
\label{sect:constraints}

One can express the final-state phase space for diffractive DIS in
terms of the transverse momentum variable, $\ptsq$, defined in the
previous section. To calculate the diffractive structure function, one
integrates over the full range of $\ptsq$. However, the large
pseudo-rapidity gap cuts imposed in some analyses of diffractive DIS
at HERA restrict the kinematically accessible range of $\ptsq$.

To see this, one calculates the boost that relates angles in the HERA
LAB and $\gamma^{*}-\pom$ CMS systems in terms of the parameters
$Q^2$, $\beta$ and $\xpom$ and the proton and electron initial
energies (see\,\cite{Ellis:1998qt} for details). Since pseudo-rapidity
is related to the LAB scattering angle, and $\ptsq$ to the CMS
scattering angle, we have the result for a pseudo-rapidity cut
$\eta_{\rm max}$:

\begin{equation}
\eta_{\rm max}\Rightarrow\theta^{\rm lab}_{\rm min}
\Rightarrow\theta^{\rm cms}_{\rm min}\Rightarrow\ptsqmin.
\end{equation}

\begin{figure}[htp]
\centering
\includegraphics[totalheight=.35\textheight,trim=70 188 100
125,clip]{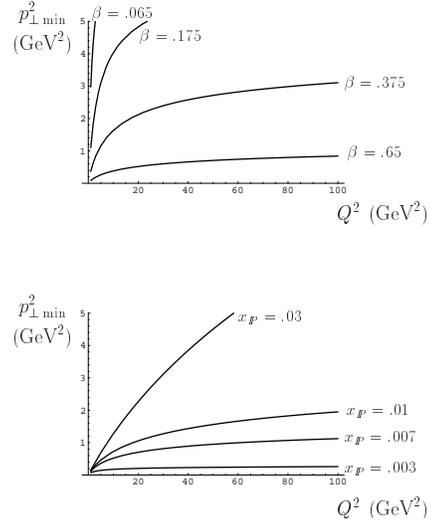}
\caption{{\it Dependence of $\ptsqmin$ on $Q^2$, $\beta$, and
 $x_{I\!\!P}$. Both graphs correspond to a pseudo-rapidity cut of
 $\eta_{\rm max}=1.8$, and for the top graph $x_{\pom}=0.007$, while
 for the lower graph $\beta=0.6$.}}\label{fig:etacut18}
\end{figure}

\begin{figure}[htp]
\centering
\includegraphics[totalheight=.35\textheight,trim=70 208 40
100,clip]{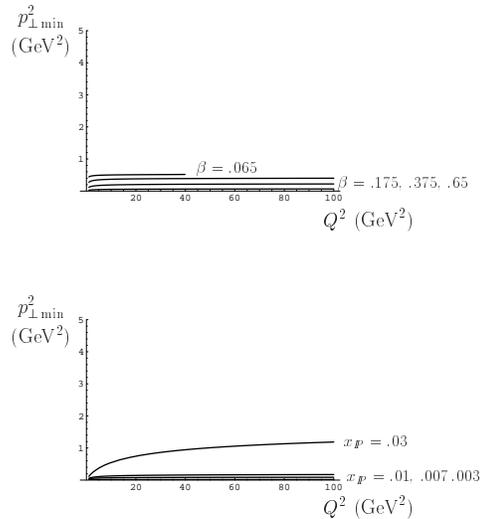}
\caption{{\it Dependence of $\ptsqmin$ on $Q^2$, $\beta$, and
$x_{I\!\!P}$, for a pseudo-rapidity cut of $\eta_{\rm max}=3.2$. For
the top graph $x_{\pom}=0.007$, while for the lower graph
$\beta=0.6$.}}\label{fig:etacut32}
\end{figure}

\noindent
In Figs.\,\ref{fig:etacut18} and\,\ref{fig:etacut32}, we show the
dependence of $\ptsqmin$ on $Q^2$, $\beta$ and $\xpom$ for dijet
production for a very strong pseudo-rapidity cut of $\eta_{\rm
max}=1.8$, corresponding to early H1 analyses\,\cite{Ahmed:1994nw},
and for the much weaker cut of $\eta_{\rm max}=3.2$ which has been
used in recent analyses\,\cite{Ahmed:1995ns,Adloff:1997nn}. It is
important to note that for a large range of $Q^2$, $\beta$ and
$\xpom$, one finds that $\ptsqmin\lsim 1\,{\rm GeV}^2$, that is, there
is no strong constraint. However, even for the weaker cut, we see that
$\ptsqmin$ can be significant at large $\xpom$, and at small $\beta$.

\subsection{Diffractive Structure Function}\label{sect:f2d3}

The diffractive structure function can be expressed as an integral
over the diffractive scattering cross section via

\begin{equation}\label{eq:f2d3lrg}
F_2^{D(3)}(Q^2,\beta,\xpom)\sim\int_{\ptsqmin}^{\frac{M_X^2}{4}}\,\rmd\ptsq\,
\frac{\rmd^4\sigma^{eP\to ePX}}{\rmd Q^2\rmd\beta\rmd\xpom\rmd\ptsq}.
\end{equation}

\noindent
From Eq.\,\ref{eq:f2d3lrg}, it is clear that for cuts which constrain
$\ptsqmin$ to be greater than a few GeV$^2$, we are not extracting the
full diffractive structure function from the data. In particular,
since $\ptsqmin$ varies with $Q^2$, $\beta$ and $\xpom$, there is a
systematic decrease in the extracted $F_2^{D(3)}$ compared to that
which would be extracted using techniques which do not impose such
constraints.

\subsection{Multi-Jet Final States} 
\label{sect:multijet}

We have also calculated the constraints for production of $q\bar{q}g$
and higher-order diffractive final states, and find a weaker
constraint applies:

\begin{equation}
\ptsqmin_{, \rm multi-jet}\approx\frac{1}{3}\ptsqmin_{, \rm dijet}\,.
\end{equation}

\noindent
Hence, in regions where $\ptsqmin_{, \rm dijet}$ is significant, one
would expect to find a relative enhancement in the contribution to the
diffractive structure function from the rapidity gap data from
multi-jet final states over that in the full diffractive structure
function, since

\begin{eqnarray} \label{eq:f2d3lrg3jets}
F_{2, \rm multi-jet}^{D(3)}(Q^2,\beta,\xpom)\sim \nonumber \\
\int_{\frac{1}{3}\ptsqmin_{, \rm dijet}}^{\frac{M_X^2}{4}}\,\rmd\ptsq\,
\frac{\rmd^4\sigma}{\rmd Q^2\rmd\beta\rmd\xpom\rmd\ptsq}\,.
\end{eqnarray}

\subsection{Factorization and pomeron structure function}

In the Ingelman-Schlein model of diffractive
DIS\,\cite{Ingelman:1985ns}, one assumes the pomeron to behave
somewhat like an hadronic state, and expects the diffractive structure
function to factorize into the product of an $\xpom$-dependent pomeron
flux factor, $f_{\pom/P}(\xpom)$, and an $\xpom$-independent ``pomeron
structure function'', $F_2^{\pom}(Q^2,\,\beta)$, via

\begin{equation} \label{eq:xpfactorization}
F_2^{D(3)}(Q^2,\beta,\xpom)=f_{\pom/P}(\xpom)F_2^{\pom}(Q^2,\beta)\,.
\end{equation}

\noindent
However, from Eq.\,\ref{eq:f2d3lrg}, we see that the LRG diffractive
structure function depends on $\ptsqmin$. Hence, since $\ptsqmin$ is a
rather sensitive function of $\xpom$, even though the full diffractive
structure function might factorize, $F_2^{D(3)}$ extracted from LRG
data for which $\ptsqmin$ is significant is not expected to factorize
due to the additional $\xpom$ dependence introduced through the lower
limit of the phase space integral. This breakdown of
$\xpom$-factorization through data selection cuts means that one
cannot extract a well-defined $\xpom$-independent pomeron structure
function in the region of parameter space in which there are phase
space restrictions due to LRG cuts.

\section{Summary}

We have discussed the result\,\cite{Ellis:1996cg} that pseudo-rapidity
cuts restrict the phase space available for diffractive deep-inelastic
scattering for some range of $Q^2$, $\beta$ and $\xpom$, by
expressing the phase space constraints in terms of a restriction on
the transverse momentum structure of the diffractive final state. For
the weaker pseudo-rapidity cuts employed in recent H1
analyses\,\cite{Ahmed:1995ns,Adloff:1997nn}, we still expect a
reduction in the diffractive structure function extracted at large
$\xpom$, and at low $\beta$ over that observed with other techniques
such as leading proton detection.

Since the phase space constraints depend on $\xpom$, this effect will
also lead to a breakdown of $\xpom$-factorization in the extracted
diffractive structure function. Further, we find that the constraints
on diffractive final states with three or more partons are rather
weaker than on dijet production\,\cite{Ellis:1998qt}, and hence also
predict that LRG cuts lead to a comparative enhancement of $q\bar{q}g$
and higher-order events over dijet events.

\vskip5mm

\noindent
{\it Acknowledgments.}  Thanks to my collaborators, Graham Ross and
John Ellis, and to the organizers of DIS99 for very interesting
conference.

\providecommand\singleletter[1]{#1}

\end{document}